\documentclass[preprint2]{aastex}
\usepackage{natbib}
\shorttitle{AzTEC 1.1 mm Observations of the MBM12 Molecular Cloud}
\shortauthors{Kim et al.}

\slugcomment{Submitted to Astrophysical Journal}
\begin{document}

\title{AzTEC 1.1 mm Observations of the MBM12 Molecular Cloud}
\author{\small M.J. Kim\altaffilmark{1}, S. Kim\altaffilmark{1}, S. Youn\altaffilmark{1},
M.S. Yun\altaffilmark{2}, G.W. Wilson\altaffilmark{2}, I. Aretxaga\altaffilmark{3}, J.P. Williams\altaffilmark{4}, D.H. Hughes\altaffilmark{3}, A. Humphrey\altaffilmark{3}, J.E. Austermann\altaffilmark{5}, T.A. Perera\altaffilmark{6},
P.D. Mauskopf\altaffilmark{7}, L. Magnani\altaffilmark{8}, and Y.-W. Kang\altaffilmark{1}}
\altaffiltext{1}{Department of Astronomy \& Space Science, Sejong
University, KwangJin-gu, KunJa-dong 98, Seoul, 143-747, Republic of Korea; sek@sejong.ac.kr}
\altaffiltext{2}{Department of Astronomy, University of Massachusetts,
710 North Pleasant St., Amherst, MA 01003, USA}
\altaffiltext{3}{Instituto Nacional de Astrof\'isca, \'Optica y Electr\'onica, Tonantzintla, Puebla, Mexico}
\altaffiltext{4}{Institute for Astronomy, University of Hawaii, 2680 Woodlawn Drive, Honolulu, HI 96822, USA}
\altaffiltext{5}{Astrophysical and Planetary Sciences, University of Colorado, Boulder, Colorado, CO 80309, USA}
\altaffiltext{6}{Department of Physics, Illinois, Wesleyan University, Bloomington, IL 61701, USA}
\altaffiltext{7}{Physics and Astronomy, Cardiff University, Wales, UK}
\altaffiltext{8}{Physics and Astronomy, The University of Georgia, Athens, GA 30602-2451, USA}

\begin{abstract}
We present 1.1 mm observations of the dust continuum emission from the MBM12 high-latitude molecular cloud observed with the Astronomical Thermal Emission Camera (AzTEC) mounted on the James Clerk Maxwell Telescope on Mauna Kea, Hawaii. We surveyed a 6.34 deg$^2$ centered on MBM12, making this the largest area that has ever been surveyed in this region with submillimeter and millimeter telescopes. Eight secure individual sources were detected with a signal-to-noise ratio of over 4.4. These eight AzTEC sources can be considered to be real astronomical objects compared to the other candidates based on calculations of the false detection rate. The distribution of the detected 1.1 mm sources or compact 1.1 mm peaks is spatially anti-correlated with that of the 100 $\mu$m emission and the $^{12}$CO emission.
We detected the 1.1 mm dust continuum emitting sources associated with two
classical T Tauri stars, LkH$\alpha$262 and LkH$\alpha$264. Observations of spectral energy distributions (SEDs) indicate that LkH$\alpha$262 is likely to be Class II (pre-main-sequence star), but there are also indications that it could be a late Class I (protostar). A flared disk and a bipolar cavity in the models of Class I sources lead to more complicated SEDs.
From the present AzTEC observations of the MBM12 region, it appears that other sources detected with AzTEC are likely to be extragalactic and located behind MBM12. Some of these have radio counterparts and their star formation rates are derived from a fit of the SEDs to the photometric evolution of galaxies in which the effects of a dusty interstellar medium have been included.
\end{abstract}
\keywords{galaxies: high-redshift --- ISM: molecules --- stars: pre-main sequence --- star: protostars --- submillimeter: galaxies --- submillimeter: stars  Online-only material: color figures}

\section{Introduction}
\label{s:intro}

The MBM12 molecular cloud was cataloged as L1453$-$L1454, L1457, and L1458
in Lynds'Dark Cloud Catalog \citep{lyn62}and is one of the nearby
high-latitude molecular clouds \citep{mbm85}. Until recently, MBM12 was considered to be the nearest star-forming region; however, its distance has been revised to 275 pc by \citet{luh01} and at 360$\pm$30 pc from the photometric and spectroscopic results
of the Sloan Digital Sky Survey (SDSS) and the Two Micron All Sky Survey (2MASS; \citet{ander02}). Thus, although it is still nearby, it is farther out than the Taurus dark clouds and Ophiuchus star-forming regions.

A few young stars were reported in MBM12 (\citet{hearty00}; \citet{hog03}) and their presence demonstrated that star formation had occurred recently in at least this particular high-latitude molecular cloud \citep{broeg06}. The high-latitude molecular clouds are known to be translucent, and the low-density environments found in high-latitude clouds create a challenge for star formation by gravitational collapse. The ability of at least a few high-latitude clouds to form cold molecular cores and young stars may be due to a combination of conditions, including variations in the interstellar radiation field, changes in dust-grain size and chemistry, and turbulence in the interstellar medium (ISM).

Regardless of the formation mechanisms, the study of high-latitude star formation environment continues to intrigue researchers, because it is
difficult to see how star formation in the low-density environments found in typical high-latitude molecular clouds can proceed via direct gravitational
collapse.
High-latitude pre-main-sequence (PMS) stars found in isolation could be formed by ejection from their parent cloud or dissipation of the natal cloud. Dissipation timescales for these unbound clouds appear to be a few Myr. It is possible that outflows from low-mass stars that form in low-density environments can drive the dissipation of the parent cloud (\citet{sd95}; \citet{fei96}).

In the present paper, we describe the results of our observations of the
dust continuum emission from MBM12 at a wavelength of 1.1 mm observed with
the AzTEC \citep{wilson08} on the James Clerk Maxwell Telescope (JCMT).
Dust continuum emission at (sub)millimeter wavelength can identify the star forming sites in the molecular clouds and determine the evolutionary stages of young stellar objects. In Section 2, we describe the observations and data reduction procedure. In Section 3, we present information on the sources detected at 1.1 mm and investigate their spectral energy distributions (SEDs).
We also describe the assembly of the SEDs and the procedure used to fit the models were assembled and how models were fitted to the data to determine the physical parameters of the detected sources from present observations. In Section 4, we summarize the results of our observations and analysis.

\section{Observations and Data Reduction}
\label{s:observe}

Observations of MBM12 were performed in raster scan mode
between 2005 November and 2006 January using AzTEC mounted on the
JCMT. The beam size was 18 arcsec, and the scan velocity was 120
arcsec s$^{-1}$. AzTEC is a new bolometric array instrument and was
developed for the Large Millimeter Telescope at the University of
Massachusetts at Amherst \citep{wilson08}.
This array camera consists of 144 Si$_3$N$_4$ micromesh bolometers and
operates at 1.1 mm and 2.1 mm. In 2005, this bolometer array was tested by invitation at the JCMT on Mauna Kea in Hawaii. It is an appropriate instrument for
surveys of Submillimeter Galaxies (SMGs) at low and high redshifts and searches for
protostellar dusty cores in nearby molecular clouds. We surveyed about a 6.34 deg$^{2}$ of the sky centered on R.A.=$02^{\textrm{h}}56^{\textrm{m}}.0$, Dec.=$19^{\textrm{d}}30^{\textrm{m}}.0$ in J2000.
This is the largest area centered on MBM12 that has ever been surveyed using continuum instruments in the submillimeter and millimeter parts of the spectrum. The area was covered with a total of 36 scan maps. Twenty seven scans were retrieved and combined, covering a total
observed area of 655.35 arcmin$^2$. The average rms was 9.695 mJy beam$^{-1}$.
AzTEC suffered a pointing error of about 2$''$. Beam maps were made
from observations of Uranus with a mean flux density of 42.2 Jy
at 1.1 mm during the JCMT observing run. A pointing model was
generated using CRL618, a post-AGB star with a mean flux density
of 2.7 Jy at 1.1 mm. The intensity calibration errors were estimated
to be $\pm$6\%--13\% \citep{wilson08}. The raw bolometer array
observations were calibrated and edited in the standard AzTEC IDL
software (\citet{scott08}; \citet{auster10}). The AzTEC instrument team
has constructed data-reduction codes in the IDL language. Using the AzTEC software,
we cleaned and co-added the array observations. To remove atmospheric signals, we adopted the principal component analysis (PCA) technique (\citet{heyer97}; \citet{francis99}), a tool for simplifying several parameters to clarify their relationships. Given a sample of $n$ objects with $p$ parameters, $x_{j}$ ($j$ = 1,..., $p$), the orthogonal variables, $\xi$$_{i}$ ($i$ = 1,..., $p$), can be described as:

\begin{equation}
\xi_{i} = a _{i1}x_{1} + \cdot\cdot\cdot + a_{ij}x_{j} + \cdot\cdot\cdot + a_{ip}x_{p}.
\end{equation}

The $\xi_{i}$ represents the principal component, and the correlated and observed variables flag important principal components \citep{brunt02}. Hence, atmospheric signals with large spatial fluctuation show dominant principal components, whereas point signals such as distant
galaxies do not. We adapted the PCA cleaning routine by constructing a covariance matrix from the $N_{\textrm{bolo}} \times N_{\textrm{time}}$ time-stream data (\citet{scott08}; \citet{auster10}). To provide an eigenvalue decomposition on the matrix, we neglected the eigenfunctions corresponding to eigenvalues larger than 2.5 $\sigma$ as atmospheric effects (\citet{scott08}; \citet{auster10}). To maximize the signal-to-noise ratio (S/N), the four cleaned raw data sets were co-added with the Wiener filter, $\it{g(q)}$, described as

\begin{equation}
g(q) = \frac{s^{\ast}(q)/J(q)}{\int|s(q)|^{2}/J(q)d^{2}q}
\end{equation}

where $\it{s(q)}$ is the Fourier transform of the beam shape from map space to spatial
frequency space, $\it{q}$, and $\it{J(q)}$ implies the average power spectral densities
(PSDs). Through filtering with the Wiener filter (\citet{scott08}; \citet{auster10}) in Fourier space, the noise at low and high frequencies was effectively
attenuated \citep{laurent05}. By using the Wiener filter and removing the PCA
components it allowed that rms noise per pixel was reduced, but the resolution of the
optimally filtered map appeared to be the same. The peak brightness of point sources and
compact peaks would be also preserved. However, after optimal filtering extended sources
in the map will be trimmed to remove areas of low coverage \citep{enoch07}. The final image of MBM12 contains eight sources detected at an S/N of greater than 4.0 $\sigma$ (Figure 1). The average rms noise over the entire image is 2.71 mJy beam$^{-1}$.
The catalog of the AzTEC mm source candidates is listed in Table 1. The flux values of the source candidates range from 10$\pm2.7$ mJy to 12$\pm2.7$ mJy.

We generated noise maps by randomly multiplying each scan in the cleaned time stream data by $\pm$1 and co-adding (\citet{perera08}; \citet{scott08}).
The sources were removed, and the random noises were accumulated in the co-added noise map \citep{auster10}. We constructed 100 noise maps by iteration.
We determined the false detection rate (FDR) by analyzing the noise properties of these maps by counting signals detected with a flux over a threshold in each noise map and then estimating the average detections \citep{perera08}. The FDR was defined as the source detections divided by the average false detections. The FDR at each threshold is shown in Figure 1. At over 3.0 $\sigma$, 312 sources were detected, and the FDR was 319$\pm$66. However, at over 4.4 $\sigma$, eight sources were detected, and the FDR value was smaller than 1. Therefore, these eight AzTEC sources are more likely to be real astronomical objects than the others. The results from the calculations of the FDR are summarized in Table 1.

\begin{figure}
\plotone{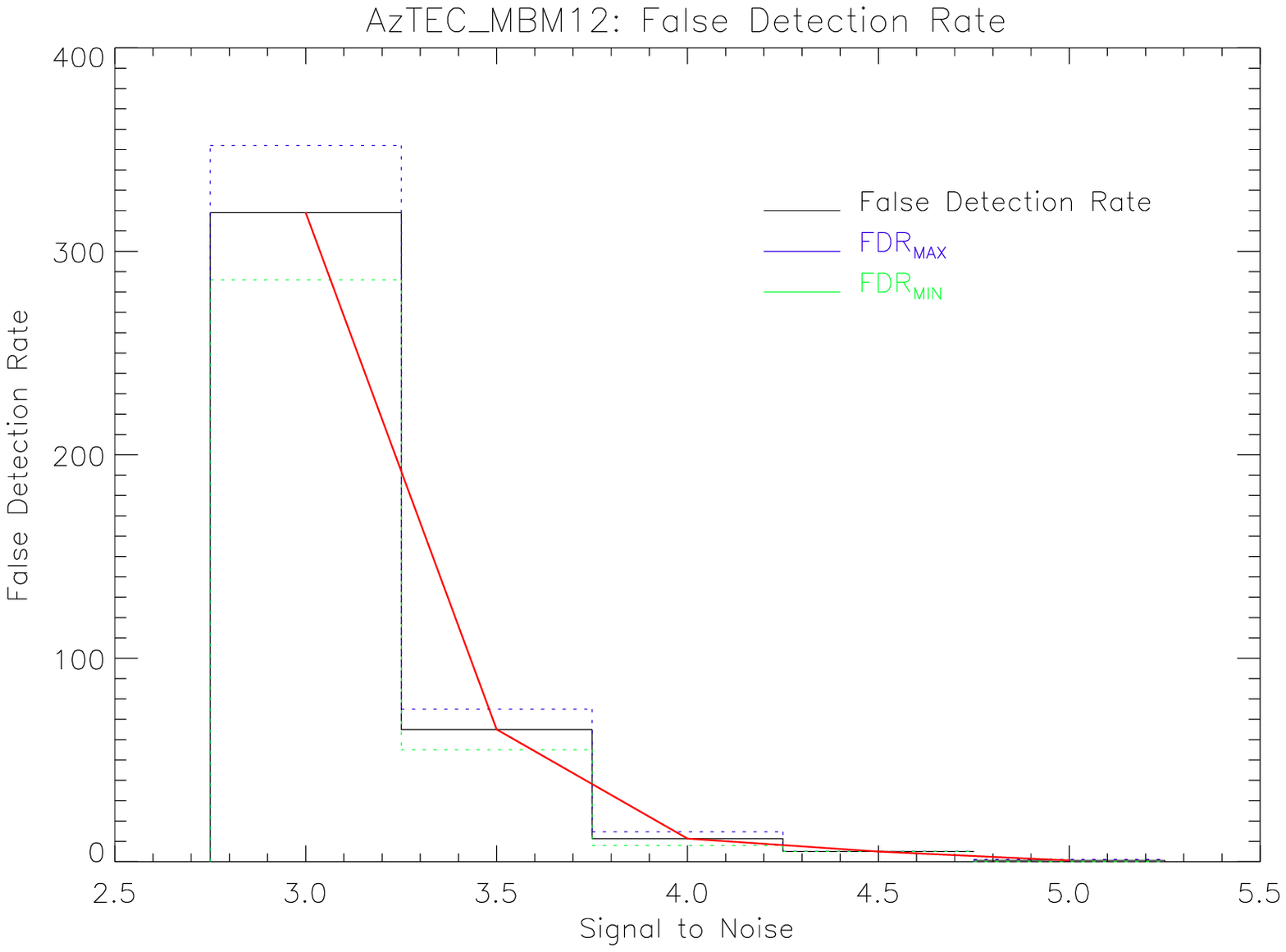}
\caption{FDR at each threshold of $3\sigma$--$5\sigma$, each
step corresponds to $0.5\sigma$. \label{fig:fdr} \newline
(A color version of this figure is available in the online journal.)}
\end{figure}

\begin{deluxetable}{cccc}
\tablewidth{0pt}
\tablecaption{The Calculated FDR of the AzTEC Sources in the MBM12 Regions. \label{table:fdr}}
\tablehead{\colhead{Area} & \colhead{Signal} & \colhead{Average Noise rms} & \colhead{FDR} \\
 & ($>5\sigma$) & (Jy beam$^{-1}$) &}
\startdata
	A1 & 1 & 0.009 & 0.18$\pm$0.10 \\
	A2 & 2 & 0.007 & 0.19$\pm$0.81 \\
	A3 & 0 & 0.007 & 0 \\
	A4 & 2 & 0.011 & 0.36$\pm$0.09 \\
	A9 & 1 & 0.013 & 0.64$\pm$0.54 \\
\enddata
\end{deluxetable}

\section{Results and Discussion}
\subsection{Distribution of the AzTEC 1.1 mm Dust Continuum Sources}

Figure 2 presents the 1.1 mm dust continuum-emitting AzTEC sources
from the present AzTEC MBM12 survey.  The $^{12}$CO(1--0) contours from the
AT\&T Bell Lab, 7-m radio telescope survey of MBM12 \citep{pound90}
are shown overlaid on the \emph{IRAS} 100 $\mu$m emission in Figure 3.  It is
clear from the figure that the $^{12}$CO contours are spatially correlated
with the 100 $\mu$m emission.  The AzTEC 1.1 mm dust continuum sources
appear to be spatially anti-correlated with the most intense $^{12}$CO-emitting
region and, thus, with the more intense 100 $\mu$m \emph{IRAS} regions in the
cloud.  The lack of AzTEC mm sources and compact peaks in the bright 100 $\mu$m emitting cores suggests that our AzTEC sources detected with the 1.1 mm bolomoter
camera traces a colder dust component in the region and the extragalactic sources.
Figure 4 shows the 1.1 mm dust continuum AzTEC sources overlaid on the images of the
100 $\mu$m emission. The anti-correlation between the AzTEC dust continuum sources and the peaks of 100 $\mu$m \emph{IRAS} and $^{12}$CO(1--0) emission is clear from the figure.

\begin{figure}
\plotone{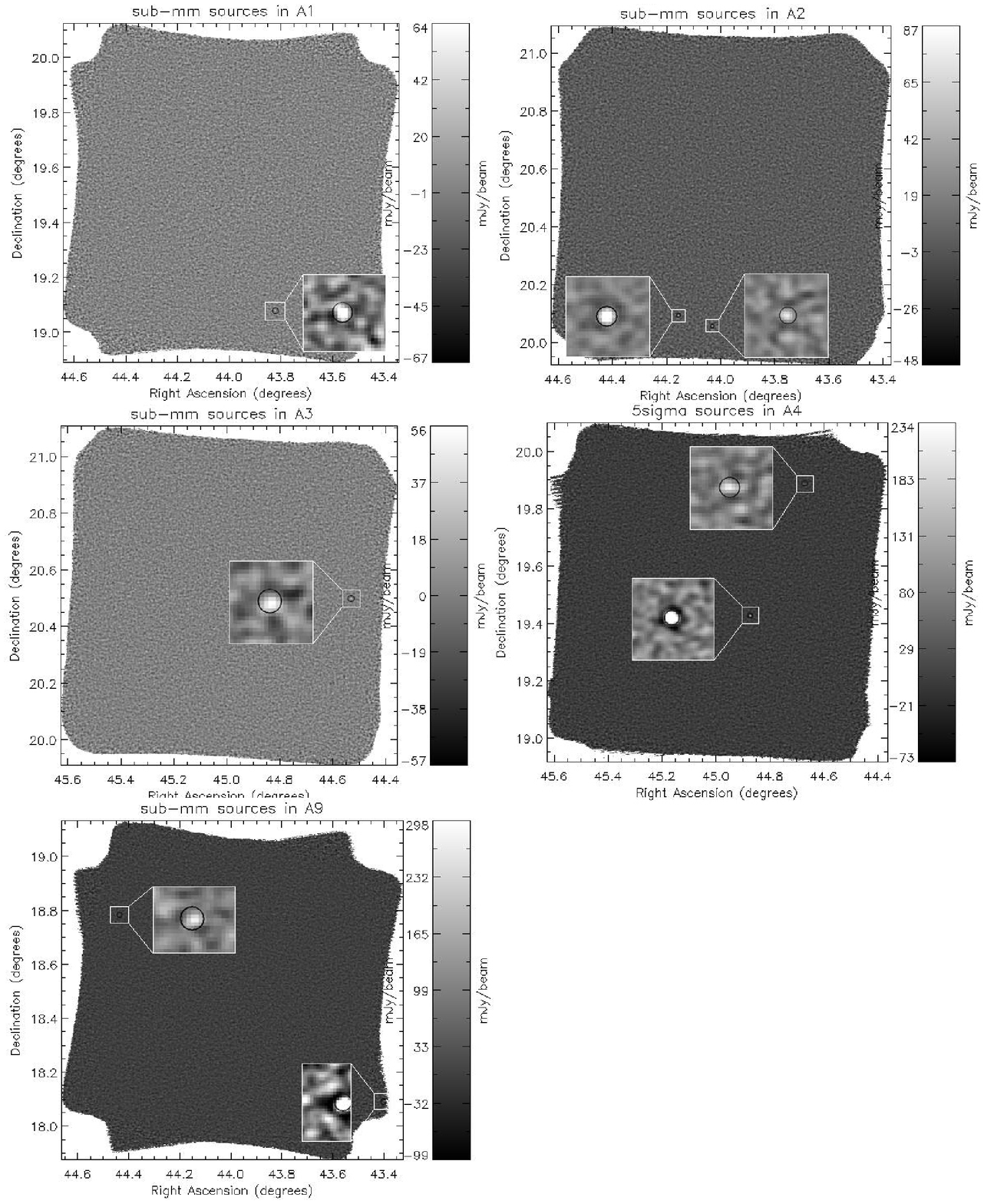}
  \caption{
  Eight AzTEC sources listed in Table 2 detected in the five regions covering the MBM12 high-latitude cloud.
  \label{fig:aztec}}
\end{figure}

\begin{figure}
    \hspace{0.1cm}\includegraphics[scale=0.35]{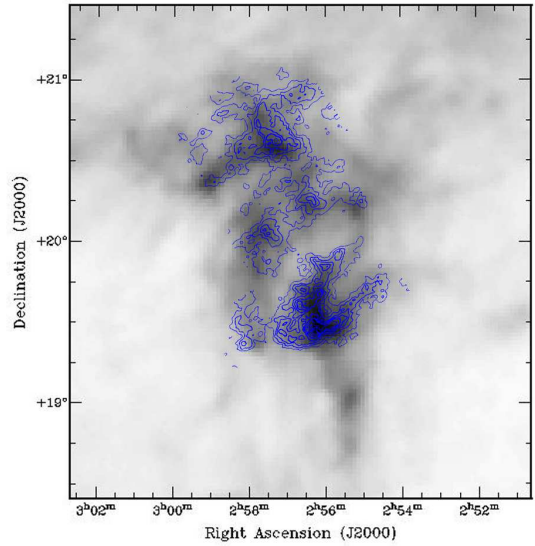}
   \caption{$^{12}$CO contours from Pound et al. (1990) are overlaid on the \emph{IRAS} 100 $\mu$m image of the MBM12.
   \label{fig:co_iras}}
\end{figure}

\begin{figure}
   \hspace{0.01cm}\includegraphics[scale=0.35]{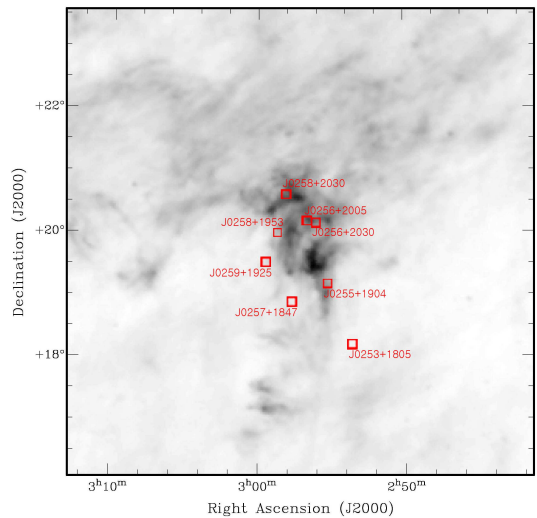}
   \caption{Eight AzTEC sources listed in Table 2 overlaid on the \emph{IRAS} 100 $\mu$m image. \label{fig:iras}}
\end{figure}


\begin{deluxetable}{cccccc}
\tablewidth{0pt}
\tablecaption{Observed AzTEC Sources in the MBM12 Region. \label{tab:source}}
\tablehead{\colhead{Source}& \colhead{ } & \colhead{R.A. (J2000)} & \colhead{Dec. (J2000)} & \colhead{Flux (mJy beam$^{-1}$)} & \colhead{S/N}}
\startdata
	A$1_1$& J0225+1904 & 02:55:17.3 & 19:04:48 &  45.6$\pm$ 3.6 &  5.2 \\
	A$2_1$& J0256+2005 (LkH$\alpha$262) & 02:56:37.5 & 20:05:36 &  87.8$\pm$ 7.0 & 12.6 \\
	A$2_2$& J0256+2003 (LkH$\alpha$264) & 02:56:07.7 & 20:03:24 &  44.9$\pm$ 3.6 &  6.3 \\
	A$3_1$& J0258+2030 & 02:58:07.3 & 20:30:00 &  29.6$\pm$ 2.4 &  4.4 \\
	A$4_1$& J0259+1925 & 02:59:29.5 & 19:25:42 & 234.3$\pm$18.7 & 22.7 \\
	A$4_2$& J0258+1953 & 02:58:41.3 & 19:53:18 &  50.7$\pm$ 4.0 &  5.0 \\
	A$9_1$& J0253+1805 & 02:53:34.8 & 18:05:42 & 298.5$\pm$23.9 & 11.4 \\
	A$9_2$& J0257+1847 & 02:57:45.6 & 18:47:06 &  53.6$\pm$ 4.2 &  4.4 \\
\enddata
\tablecomments{Units of R.A. are in hours, minutes, and seconds. Units of Dec. are in degrees, arcminutes, and arcseconds. The positional uncertainties of sources are likely to be an order of 1$''$ in R.A. and Dec.}
\end{deluxetable}

\subsection{Properties of the AzTEC Sources in the MBM12}

We present eight AzTEC millimeter sources in Figure 2. These sources are overlaid on the \emph{IRAS} 100 $\mu$m image in Figure 3. The catalog of detected sources
is presented in Table 2. The sources in the catalog were detected with an S/N of over
4.4$\sigma$. Each point source is indicated on the co-added map of the MBM 12 region. The size of the
circle in the box is 18 arcsec which is the beam size of the AzTEC
observations for the present survey. To identify the nature of these sources and their properties, we examined their SEDs. The intrinsic flux densities of the detected AzTEC sources were used for the analysis of their SEDs \citep{perera08}. Among the eight AzTEC sources, two sources
(A$2_1$, A$2_2$) in Table 1 correspond to known T Tauri stars, LkH$\alpha$ 262 and LkH$\alpha$ 264 with strong H$\alpha$ emission and Li I absorption (\citet{hearty00}; \citet{luh01}). Studies of seven T Tauri stars in this region were conducted previously (Table 3).

\subsubsection{AzTEC Sources with Lower FDR}

\emph{J0256+2005 (A$2_1$) and J0256+2003 (A$2_2$)}: we detected 1.1 mm
dust continuum emitting sources associated with LkH$\alpha$ 262 (J0256+2005) and LkH$\alpha$ 264 (J0256+2003). Previous studies by \citet{hog03}
 reported observing continuum emission from the disks around the classical T Tauri stars, LkH$\alpha$ 262, LkH$\alpha$ 263, and LkH$\alpha$
264 at 450 $\mu$m and 850 $\mu$m with Submillimeter Common User
Bolometer Array (SCUBA). These two stars, LkH$\alpha$ 262 and LkH$\alpha$ 264, emitting submillimeter and millimeter radiation, were reported as Class II (\citeauthor{hea99}\citeyear{hea99}; \citeauthor{luh01}\citeyear{luh01}; \citeauthor{hog03}\citeyear{hog03}). The SEDs in Figure 5
indicate that LkH$\alpha$ 262 is likely to be Class II, but there are also indications that this source might be a late Class I (\citeauthor{and93}\citeyear{and93}; \citeauthor{whiteny03}\citeyear{whiteny03}) instead. The flux flattens towards the longer wavelengths, a phenomenon that can be seen in the late Class I.
To provide a reasonable fit to the flux densities between 3.6 and 8.0 $\mu$m for the SED of LkH$\alpha$ 262, we used a one dimensional radiative transfer model of DUSTY \citep{ivezic99}.
The physical properties of the central source and the dusty envelope were
derived from the modeling. The flux densities from the central source were absorbed by the dusty envelope and re-emitted to longer wavelengths.
A dusty disk was considered in order to fit 24 $\mu$m and 70 $\mu$m wavelengths, since it was difficult to fit the flux densities using the stellar component.
A power law density distribution for the envelope was also tested for emission from the disk model. The estimated mass of the envelope was consistent with the mass of the envelope determined from 1.1 mm continuum by fixing most of free parameters. The mass of the envelope was determined from the flux density $S$ at 1.1 mm, the Planck function at temperature $T$, and the dust opacity $\kappa$. The dust opacities used in this analysis were from OH5 dust in \citet{ossenkopf94}. The estimated mass was about 0.04 $M_\sun$ in the 15--20 K temperature range. Our detections might suggest that at least the disks of the classical T Tauri stars in MBM12 (1--5 Myr) have a mass range similar to the mass range of the disks of the younger ($\sim$1 Myr) Taurus and $\rho$ Ophiuchus regions (0.001--0.3 $M_\odot$; \citeauthor{and07}\citeyear{and07}; \citeauthor{williams11}\citeyear{williams11}). \citet{luh01} argued that the near-IR K- and L- band excesses indicated that significant disk dispersal had occurred in MBM12.

\begin{figure*}
\includegraphics[scale=.80]{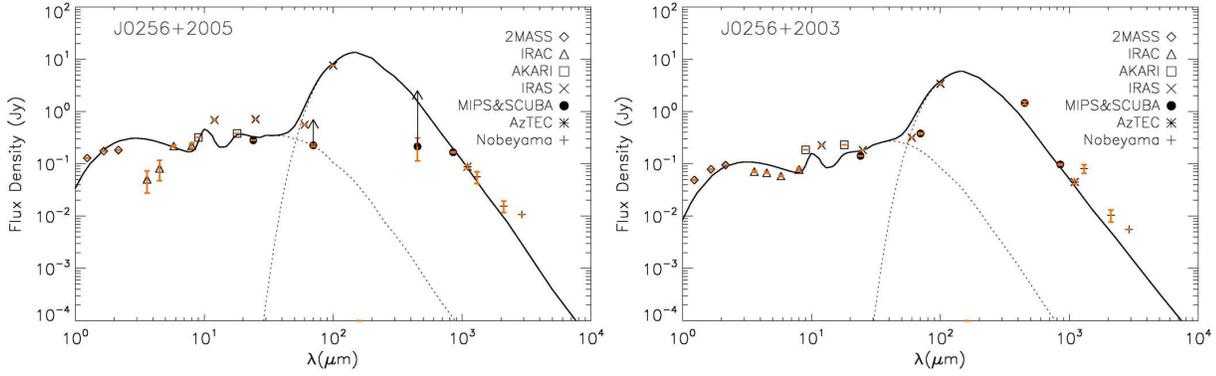}
\caption{SEDs of two AzTEC sources corresponding to the candidates of protostars, J0256$+$2005 (Left) and J0256$+$2003 (Right).\label{fig:sed1} \newline
(A color vertion of this figure is available in the online journal.)}
\end{figure*}

\begin{figure*}
\includegraphics[angle=90,scale=.63]{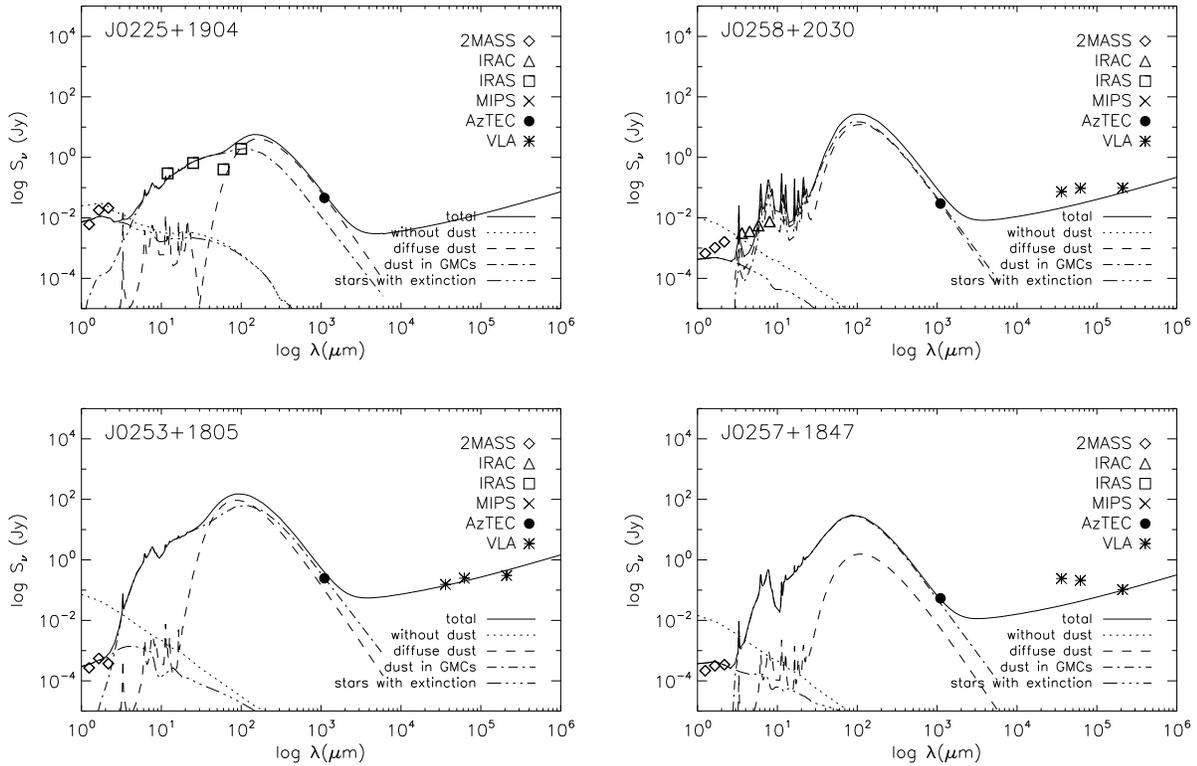}
\caption{SEDs of four AzTEC sources corresponding to the candidates of extragalactic and submillimeter galaxies, J0225+1904 (Top Left), J0258+2030 (Top Right), J0253+1805 (Bottom Left), and J0257+1847 (Bottom Right).}
\end{figure*}

\subsubsection{AzTEC Sources with Higher FDR}

\emph{RXJ0255.4+2005}: this object was classified as a weak-line T Tauri star (WTTS). Its spectral
type was labeled as K6 by \citet{hea00}. $T_{\textrm{eff}}$ was reported
as 4205 K by \citet{luh99} and 4275 K by \citet{hea00}. \citet{hog02} calculated the disk mass and diameter using the Owens Valley Radio Observatory (OVRO) to observe the $\lambda$=3 mm continuum. The results were reported as 0.04 $M_{\odot}$ and 52 AU, respectively.
These values were based on a distance of 275 pc and a gas-to-dust ratio of 100. The disk mass could be 0.023 $M_{\odot}$ at the distance of 360 pc.
There were also studies by \citet{hea00} and \citet{her04} for the line widths of H$\alpha$ from the source. The emission line widths were $-$1.26 \AA\ \citep{hea00} and $-$1.1 \AA\ \citep{her04}. The flux of this object varies, and the period was reported as 6.22 days by \citet{her04} and 3.36$\pm$0.17 days by \citet{broeg06}.

\emph{E02553+2018}: this object appears on the border between a classical and a WTTS. The emission line width of H$\alpha$ is about $-$1.6 \AA\ \citep{hea00} and $-$2.5 \AA\ \citep{her04}. Its spectral type is K3 \citep{fleming89} or K4 \citep{hea00}. The effective temperature, $T_{\textrm{eff}}$, of this object is 4660 K, as calculated by \citet{luh99} and 4593 K by \citet{hea00}. \citet{hog02} obtained a disk mass of 0.07 $M_{\odot}$ and a diameter of 66 AU. The flux was measured as 11 mJy with OVRO. The disk mass could be 0.04 $M_{\odot}$ at the distance of 360 pc. This object has an additional component of $V_{\textrm{LSR}}$=$-$1.5 km$s^{-1}$.


\emph{RXJ0258.3+1947}: this object was classified as a classical T Tauri star (CTTS). Its spectral type was determined as M5 \citep{hea00} or M4.5 \citep{luh01}. The effective temperature, $T_{\textrm{eff}}$, was calculated as 3198 K by \citet{luh99} and 3169 K by \citet{hea00}. The flux at 3 mm was 5 mJy, and the disk mass was 0.05 $M_{\odot}$. The disk mass could be 0.03 $M_{\odot}$ at the distance of 360 pc.
Its diameter was 58 AU, from OVRO observations, and JCMT observations yielded a mass of $8 \times 10^{-4}M_{\odot}$ \citep{hog02}. Its flux varies, and the period was reported as 1.205 days \citep{her04}. This was confirmed with reports of an H$\alpha$ emission line width of $-$24.5 \AA\ from \citet{hea00} and $-$25$\sim$$-$34 \AA\ from \citet{her04}.

\emph{RXJ0256.3+2005}: this object was classified as a WTTS and is the latest spectral type in the field of MBM12. Its spectral type was designated as M5.75 \citep{luh01}
and M6 \citep{her04}. \citet{luh99} calculated that the effective temperature, $T_{\textrm{eff}}$, was 3024 K. The mass was between 0.15 and 0.1 $M_{\odot}$ \citep{mee09}. Its H$\alpha$ emission line width was $-$13.5 \AA\ \citep{her04}.



\begin{deluxetable}{cccc}
\tablewidth{0pt}
\tablecaption{Previously Observed T Tauri Stars in the MBM12 \label{tab:previously}}
\tablehead{\colhead{star} & \colhead{R.A. (J2000)} & \colhead{Dec. (J2000)} & \colhead{Ref.}}
\startdata
	RXJ0255.4+2005   & 02:55:25.7 & 20:04:53 & 6  \\
	LkH$\alpha$ 262  & 02:56:07.9 & 20:03:25 & 1,2 \\
	LkH$\alpha$ 263  & 02:56:08.4 & 20:03:39 & 1,2 \\
	LkH$\alpha$ 264  & 02:56:37.5 & 20:05:38 & 1,2,3,4 \\
	E02553+2018      & 02:58:11.2 & 20:30:04 & 2,5 \\
	RXJ0258.3+1947   & 02:58:15.9 & 19:47:17 & 6 \\
	RXJ0256.3+2005   & 02:56:17.9 & 20:06:10 & 6 \\
\enddata
\tablecomments{Units of R.A. are in hours, minutes, and seconds. Units of Dec. are in degrees, arcminutes, and arcseconds.}
\tablerefs{(1) Herbig \& Bell (1988);
(2) Fern\`andez et al. (1995);
(3) Magnani et al. (1995);
(4) Gameiro et al. (1993);
(5) Caillault et al. (1995);
(6) Hearty et al. (2000)}
\end{deluxetable}


\subsection{Properties of the AzTEC Sources behind the MBM12}

From the present AzTEC observations of the MBM12 region, we find
that other detected sources are likely to be located behind MBM12. Four
AzTEC objects, defined as J0258+2030 (A$3_1$), J0259+1925 (A$4_1$), J0253+1805 (A$9_1$), and J0257+1847 (A$9_2$), were
previously studied in the 1.4 GHz NRAO VLA survey for the sky of
$\delta \ge -40^{\circ}$ in J2000 \citep{con98}. J0259+1925 (A$4_1$) is known
as quasi-stellar object (QSO) at a redshift of $z$ = 0.54. The flux
density of this object at 1.4 GHz was 169.9 mJy. Three AzTEC
sources, J0258+2030 (A$3_1$), J0253+1805 (A$9_1$), and J0257+1847 (A$9_2$) have radio counterparts observed by
\citet{con98} with the VLA at 1.4 GHz. The counterpart
corresponding to J0259+1925 (A$4_1$) was also classified as
J025929+1925.7 (1REX) in the radio-emitting X-ray sources (REXs)
survey by \citet{cac00}. Its X-ray flux was reported as
$2.34 \times 10^{-13}\, \textrm{erg}\,\textrm{s}^{-1}\,\textrm{cm}^{-2}$. The
NRAO VLA Sky Survey (NVSS) position of this source suggests that this object
should be $\alpha=02^{\textrm{h}}59^{\textrm{m}}29^{\textrm{s}}.65$, $\delta=+19^{\circ}25^{\prime}44^{\prime\prime}.9$ (J2000) at radio
wavelengths.
From the present AzTEC observations, the position of J0259+1925 was measured as $\alpha=02^{\textrm{h}}59^{\textrm{m}}29^{\textrm{s}}.5$, $\delta=+19^{\circ}25^{\prime}42^{\prime\prime}$ at millimeter wavelengths. So the difference between the radio and millimeter emission is $\Delta\alpha=0^{\textrm{s}}.15$ and $\Delta\delta=2^{\prime\prime}.9$. The flux at 1.1mm, measured with AzTEC was also measured as 234.3 mJy beam$^{-1}$. The AzTEC source A$9_1$ corresponds to J0253+1805 (J2000) in the NRAO VLA sky survey at 1.4 GHz \citep{con98}. The coordinate of J0253+1805 (A$9_1$) was reported as $\alpha=02^{\textrm{h}}53^{\textrm{m}}34^{\textrm{s}}.88$ and $\delta=+18^{\circ}05^{\prime}42^{\prime\prime}.53$ (J2000) using the VLA observations. This position agrees with the identification by the Automated Plate Measurement (APM) Facility at Cambridge of the radio sources from the Jodrell Bank VLA Astrometric Survey (JVAS) \citep{sne02}. We also compared the fluxes at 6 mm (Green Bank), NVSS 1.4 GHz (NRAO VLA Sky Survey), Green Bank 1.4 GHz, and VLA 1.4 GHz. Their results were respectively, 248, 304, 426, and 154 mJy. The millimeter-wavelength flux from this study was 298.5 mJy beam$^{-1}$ at 1.1 mm. The AzTEC source detected as J0257+1847 (A$9_2$) has its counterpart in the radio source in the NRAO VLA sky survey by \citet{con98}. This source was also included in the Very Long Baseline Array (VLBA) Calibrator Survey (VCS1) at 2.3 and 8.4 GHz \citep{bea02}. \citet{sne02} also obtained the flux densities of this object at 6 mm (Green Bank), 1.4 GHz (NVSS), 1.4 GHz (Green Bank), and 1.4 GHz (VLA), which were 209, 103, 180, and 241 mJy, respectively. Synchrotron properties and their spectral indices of these objects will be discussed in detail in a future paper (S. Kim et al., in preparation). The coordinates of the AzTEC source agrees to within $\alpha=02^{\textrm{h}}57^{\textrm{m}}45^{\textrm{s}}.6$, $\delta=+18^{\circ}47^{\prime}06^{\prime\prime}$ (J2000). The flux at 1.1 mm is 53.6 mJy beam$^{-1}$. The fluxes at other wavelengths were compared in the SEDs presented in Figure 6. Although two AzTEC sources, J0225+1904 (A$1_1$) and J0258+1953 (A$4_2$), do not have counterparts at other wavelengths, we also performed a stacking analysis \citep{scott08} on these AzTEC detections to calculate upper limit of the fluxes at the near-infrared images (Table 4). The GRASIL model \citep{silva98} computes the photometric evolution of galaxies based on the SEDs of the AzTEC sources, including the effects of a dusty interstellar medium. The model includes the escape time of young stars from molecular clouds, the fraction of residual gas in molecular clouds, the radius of molecular clouds, etc.
The star formation rate (SFR) could be also derived by fitting the SEDs to the model \citep{kim10}. The results suggest that J0258+2030 (A$3_1$), J0253+1805 (A$9_1$), and J0257+1847 (A$9_2$) can have SFRs of 3.6 $\times$ 10$^3$ $M_\sun$ yr$^{-1}$, 3.3 $\times$ 10$^4$ $M_\sun$ yr$^{-1}$, and 2.9 $\times$ 10$^2$ $M_\sun$ yr$^{-1}$, respectively. The SFRs are similar to those studied in 16 radio galaxies with AzTEC counterparts in the far-IR and a gray-body emission template with an emissivity index, $\beta$, and a dust temperature, $T_d$ \citep{hump11}.

\begin{deluxetable}{ccccccc}
\tablewidth{0pt}
\tablecaption{Fluxes at Other Wavelengths for the Detected AzTEC Sources \label{table:flux}}
\tablehead{
\colhead{source} & \colhead{2MASS J(1.25$\mu$m)} & \colhead{2MASS H(1.65$\mu$m)} & \colhead{2MASS K(2.17$\mu$m)} & \colhead{$S_{1.4\textrm{GHz}}$} & \colhead{$z$} & \colhead{Ref.} \\
 & \colhead{(mJy)} & \colhead{(mJy)} & \colhead{(mJy)} & \colhead{(mJy)} & &}
\startdata
J0255+1904 & ($<$) 6.05$\pm$0.097  & ($<$) 18.0$\pm$0.306   & ($<$) 21.3$\pm$0.341 &               & & \\
J0256+2005 & 129.0$\pm$2.064   & 175.0$\pm$2.975  & 183.0$\pm$1.994  &               & & 1,2,6,8 \\
J0256+2003 & 49.0$\pm$0.784    & 78.5$\pm$1.335 & 94.6$\pm$1.031 &               & & 1,5,6,7 \\
J0258+2030 & 0.663$\pm$0.097 & 1.04$\pm$0.067 & 1.62$\pm$0.08  & 98.1$\pm$3.0  & & 2,3 \\
J0259+1925 & ($<$) 6.05$\pm$0.097  & ($<$) 18.0$\pm$0.306   & ($<$) 21.3$\pm$0.341 & 163.7$\pm$4.9 & 0.54 & 2,9 \\
J0258+1953 & ($<$) 6.05$\pm$0.097  & ($<$) 18.0$\pm$0.306   & ($<$) 21.3$\pm$0.341 & 98.1$\pm$3.0  &&  \\
J0253+1805 & ($<$) 6.05$\pm$0.097  & ($<$) 18.0$\pm$0.306   & ($<$) 21.3$\pm$0.341 & 304.0$\pm$9.1   & 0.427 & 2,4,10 \\
J0257+1847 & ($<$) 6.05$\pm$0.097  & ($<$) 18.0$\pm$0.306   & ($<$) 21.3$\pm$0.341 & 103.7$\pm$3.1 & & 2,3,4,11 \\
\enddata
\tablerefs{(1) Cutri et al. (2003);
(2) Condon et al. (1998);
(3) Gregory et al. (1996);
(4) Snellen et al. (2002);
(5) Herbig et al. (1988);
(6) Fern\`andez et al. (1995);
(7) Magnani et al. (1995);
(8) Gameiro et al. (1993);
(9) Caccianiga et al. (2000);
(10) Ma et al. (1998);
(11) Beasley et al. (2002)}
\end{deluxetable}

\section{Summary and Conclusion}

We presented observations of the dust continuum emission at 1.1 mm from MBM12 performed with AzTEC at the JCMT. We surveyed about 6.34 deg$^2$ field of sky. This is the largest area that has ever been surveyed for the MBM12 region with a (sub)-millimeter telescope. The distribution of 1.1 mm AzTEC sources is spatially anti-correlated with that of the 100 $\mu$m emission and that of the $^{12}$CO emission. We detected 312 sources with an S/N of over 3 $\sigma$. The FDR was 319$\pm$66. Eight sources were detected and their FDR value was smaller than 1. These eight AzTEC sources can be considered to be real astronomical objects as opposed to spurious detections.
We detected 1.1 mm sources associated with two classical T Tauri stars, LkH$\alpha$262 and LkH$\alpha$264. Observations of their SEDs indicate that LkH$\alpha$262 is likely to be Class II but may be a late Class I. A flared disk and a bipolar cavity in models of Class I sources lead to more complicated SEDs. Our detections suggest that at least the classical T Tauri stars in MBM12 have disks in a similar mass range with the disks of stars in the Taurus or $\rho$ Ophiuchus regions. From the present AzTEC observations of the MBM12 region, we find that some of the other detected sources are likely to be located behind MBM12 molecular cloud. These have radio counterparts, and their SFRs were derived from fits of their SEDs to the photometric evolution of galaxies when the effects of a dusty interstellar medium were included. These SFRs ranged from a few hundred to a few thousand $M_\odot$ yr$^{-1}$. Direct measurements of their red-shifts should be performed with the Large Millimeter Telescope in the near future.

\vspace*{1cm}

We thank K. Scott, J. Lowenthal, and other AzTEC team members for their contributions to the project. We also thank S. Johnson for helpful suggestions on the manuscript. We thank the anonymous referee for careful reading and helpful comments on the manuscript. This work has made use of the NASA/IPAC Extragalactic Database (NED) which is operated by the Jet Propulsion Laboratory, California Institute of Technology, under contract with the National Aeronautics and Space Administration. This work also has made use of data products from the Two Micron All Sky Survey which is a joint project of the University of Massachusetts and the Infrared Processing and Analysis Center/California Institute of Technology and the NASA/IPAC Infrared Science Archives which is operated by the Jet Propulsion Laboratory and California Institute of Technology.
The James Clerk Maxwell Telescope is operated by the Joint Astronomy Center on behalf of SRC of the United Kingdom, the Netherlands OSR, and the Canada NRC. MK was supported in part by Basic Science Research Program through the National Research Foundation of Korea (NRF) funded by the Ministry of Education, Science and Technology 2009-0066892. This research was supported in part by Mid-career Researcher Program through the National Research Foundation of Korea (NRF) funded by the Ministry of Education, Science and Technology 2011-0028001.

\end{document}